\documentclass[12pt]{article}
\usepackage{graphics}
\usepackage{amssymb}
\usepackage{amsmath}
\usepackage{upgreek}

\begin{document}

{\noindent \bf \Large Turbulence Accelerating Cosmology from an Inhomogeneous Dark Fluid}

\bigskip

\begin{center}

I. Brevik\footnote{iver.h.brevik@ntnu.no}

\bigskip
Department of Energy and Process Engineering, Norwegian University of Science and Technology, N-7491 Trondheim, Norway

\bigskip
A. V. Timoshkin

\bigskip

Tomsk State Pedagogical University, 634061 Tomsk, Russia

\bigskip

Ye. Rabochaya

\bigskip
Eurasian National University, Astana, Kazakhstan

\bigskip

S. Zerbini\footnote{zerbini@science.unitn.it}

\bigskip

Department of Physics, University of Trento, and Gruppo Collegato di Trento, Sezione INFN di Padova, Italy

\today

\end{center}

\bigskip
\begin{abstract}

Specific dark energy models with a linear inhomogeneous time-dependent equation of state, within the framework of 4d Friedman-Robertson-Walker (FRW) cosmology, are investigated. It is demonstrated that such 4d inhomogeneous fluid models may lead to a turbulence FRW cosmology. Both one-component and two-component models from 4d inhomogeneous dark fluid models are considered. In the one-component model the universe may develop from a viscous era with, for instance, a constant bulk viscosity, into a turbulent era. In the two-component model the fluid can be decomposed into two components, one non-turbulent (ideal) and another turbulent part, obeying two different equations of state. Conditions for the appearance of the turbulent dark energy universe in terms of the parameters in the equation of state (EoS) without introducing the turbulence concept explicitly are are obtained. An equivalent description in terms of an inhomogeneous fluid for the viscous Little Rip (LR) cosmology is also developed.

\end{abstract}

\section{ Introduction}

A variety of complicated problems in cosmology can be explained by the discovery of the accelerated expansion of the universe \cite{riess98,perlmutter99} in terms of dark energy \cite{bamba12,li99,nojiri11}. According to recent observations the dark energy currently accounts for about 69\% of the total mass/energy of the universe \cite{plank}. It possesses a negative pressure and/or negative entropy. The EoS parameter $w$ is still determined up to some uncertainty: it is not clear if $w$ is less than $-1$, equal to $-1$, or larger than $-1$. According to present observations, $w=-1^{+0.09}_{-0.10}$ \cite {nakamura10,amanullah10}.

The most interesting case is when the thermodynamic parameter $w=p/\rho <-1$ (phantom dark energy). An essential property of this kind of energy is the Big Rip future singularity \cite{caldwell02} (see also \cite{nojiri03,nojiri05}), where the scale factor becomes infinite at a finite time in the future. In the mild phantom models where $w$ asymptotically tends to $-1$, the singularity occurs in the infinite future \cite{frampton11,frampton12,astashenok12,frampton11B}. Such Rip phenomena take place for mild phantom scenarios like Little Rip or Pseudo Rip.

In a series of previous works \cite{brevik12,brevik13,rabochaya13,brevik11} we considered the non-viscous models of the cosmic fluid. The case of such a fluid (also called an ideal fluid) is quite an idealized model; it will often be useless in practical situations, especially when fluid motion near boundaries is involved. Also under boundary-free conditions (isotropic turbulence, for instance), the influence from viscosity can be most important.

When working to the first order in deviations from thermodynamic equilibrium one has in principle to introduce two viscosity coefficients, namely the shear viscosity $\eta$ and the bulk viscosity $\zeta$. We shall assume, in conformity with usual practice, that spatial anisotropies (present in in the Kasner universe, for example), become smoothed out. Thus only the coefficient $\zeta$ will be included.

\bigskip

In the present article we point out the equivalence between

1) expansion of the universe described in terms of time-dependent parameters of the inhomogeneous dark fluid model, and

2) viscous Little Rip (LR) cosmology for the dark fluid in the late universe.

\bigskip

Our work is based upon, and extends, prior work of Ref.~\cite{brevik12B} (see also Refs.~\cite{Houndjo:2012be,gron13}).

\section{Dark fluid with bulk viscosity}

A theory of viscous LR cosmology was recently given in Ref.~\cite{brevik11}. We consider now viscous LR cosmology in an isotropic cosmic fluid in the later stages of the development of the universe.

We shall assume viscosity-dependent governing equations. We suppose that the viscosity function $\xi(H)$, defined as $3\zeta H$, is a constant:
\begin{equation}
3\zeta H \equiv \xi_0=\rm const, \label{1}
\end{equation}
with $H=\dot{a}/a$ the Hubble parameter.
Then the expression for the time dependent energy density becomes \cite{brevik11}
\begin{equation}
\rho(t)=\left[ \left(\frac{\xi_0}{A}+\sqrt{\rho_0}\right)\exp (\sqrt{6\pi G}\,At)-\frac{\xi_0}{A}\right]^2, \label{2}
\end{equation}
with $A$ a positive constant. This is a characteristic property of LR cosmology, now met under viscous conditions.

Next, let us consider LR cosmology from the point of view of 4d FRW non-viscous cosmology. Here it is natural to associate $t=0$ with the present time, so that $\rho_0$ becomes the present time energy density. The Friedman equation for a spatially flat universe is
\begin{equation}
\rho=\frac{3}{k^2}H^2, \label{3}
\end{equation}
where $\rho$ is the energy density, and $k^2=8\pi G$.

Assume that our universe is filled with an ideal fluid (dark energy) obeying an inhomogeneous equation of state \cite{brevik04,brevik13}
\begin{equation}
p=w(t)\rho+\Lambda(t), \label{4}
\end{equation}
with $w(t)$ and $\Lambda(t)$ as time-dependent parameters and $p$ the pressure.

The energy conservation law is
\begin{equation}
\dot{\rho}+3H(p+\rho)=0, \label{5}
\end{equation}
and the derivative of $\rho$ with respect to cosmic time is
\begin{equation}
\dot{\rho}=6\sqrt{2\pi G}(\xi_0+A\sqrt{\rho_0})\exp (\sqrt{6\pi G}\,At)H. \label{6}
\end{equation}
Taking into account Eqs.~(\ref{2}), (\ref{4})-(\ref{6}) we obtain
\begin{equation}
2\sqrt{2\pi G}(\sqrt{3}\, AH+\xi_0)+[1+w(t)]3AH^2+\Lambda(t)=0. \label{7}
\end{equation}
Solving with respect to $\Lambda(t)$, we have
\begin{equation}
\Lambda(t)=-3A[1+w(t)]H^2-2\sqrt{2\pi G}(\sqrt{3}\, AH+\xi_0). \label{8}
\end{equation}
If the parameter $w(t)$ is chosen as
\begin{equation}
w(t)=-1-\frac{\delta}{3AH^2}, \label{9}
\end{equation}
with $\delta$ is a positive constant, the "cosmological constant" becomes
\begin{equation}
\Lambda(t)=\delta -2\sqrt{2\pi G}(\sqrt{3}\,AH+\xi_0). \label{10}
\end{equation}
Consequently, we have achieved the solution (\ref{10}) which is in conformity with the LR expression (\ref{2}), valid when the condition (\ref{1}) is satisfied. The noticeable point is that such a LR behavior is now induced purely via the $\Lambda-$ sector.

\section{The turbulent approach}

Let us consider the dark energy universe in its later stages, where it approaches the future singularity. The fluid system may then be regarded as quasi-stationary, and it becomes natural to take into account a transition into turbulence motion.

We write the effective energy density as a sum of two terms \cite{brevik12B}:
\begin{equation}
\rho_{\rm eff}=\rho+\rho_{\rm turb}, \label{11}
\end{equation}
where $\rho$ denotes the laminar ordinary energy density and $\rho_{\rm turb}$ its turbulent part. We assume that $\rho_{\rm turb}$ is proportional to the scalar expansion $\theta={U^\mu}_{;\mu}=3H$ and write the effective energy density as
\begin{equation}
\rho_{\rm eff}=\rho(1+3\tau H), \label{12}
\end{equation}
with $\tau$ a proportionality factor.

Analogously we split the effective pressure $p_{\rm eff}$ into two terms,
\begin{equation}
p_{\rm eff}=p+p_{\rm turb}. \label{13}
\end{equation}
The non-turbulent quantities $p$ and $\rho$ are connected by the standard relationship
\begin{equation}
p=w\rho, \label{14}
\end{equation}
where $-1<w<-1/3$ in the quintessence region and $w<-1$ in the phantom region.

We take the dependence of $p_{\rm turb}$ on $\rho_{\rm turb}$ to be as simple as possible,
\begin{equation}
p_{\rm turb}=w_{\rm turb} \,\rho_{\rm turb}, \label{15}
\end{equation}
with $w_{\rm turb}$ a constant.

We shall consider two different possibilities for the value of $w_{\rm turb}$. First, we put $w_{\rm turb}$ equal to $w$ in Eq.~(\ref{14}), meaning that turbulent matter behaves in the same way non-turbulent matter as far as the equation of state is concerned. As second option, we shall assume $w_{\rm turb}$ to be different from $w$.

\bigskip

{\noindent 1) \bf Assume that $w_{\rm turb}=w<-1$.}

\bigskip

Even in this case the time development of $\rho$ will be different from that of $\rho_{\rm turb}$. The ratio between turbulent and non-turbulent energy density becomes \cite{brevik12B}
\begin{equation}
\frac{\rho}{\rho_{\rm turb}}=3\tau H=\frac{3\tau H_0}{Z}. \label{16}
\end{equation}
The Hubble parameter becomes
\begin{equation}
H=\frac{H_0}{Z}, \label{17}
\end{equation}
where we have defined $Z$ as
\begin{equation}
Z=1+\frac{3}{2}\gamma H_0 t, \label{18}
\end{equation}
with
\begin{equation}
\gamma=1+w. \label{19}
\end{equation}
Here $H_0$ is the initial value of $H$ at the present time $t=t_0$.

From the first Friedman equation we get a Big Rip behavior for the turbulent energy density \cite{brevik12B},
\begin{equation}
\rho=\frac{3H_0^2}{k^2}\,\frac{1}{Z}\,\frac{1}{Z+3\tau H_0}. \label{20}
\end{equation}
We now find
\begin{equation}
\dot{\rho}=-\frac{9\gamma H_0^3}{2k^2}\frac{2Z+3\tau H_0}{[Z(Z+3\tau H_0)]^2}. \label{21}
\end{equation}
using Eqs.~(\ref{4}), (\ref{20}) and (\ref{21}) the energy conservation can be rewritten as
\begin{equation}
\frac{3\gamma H_0^2}{Z^2}\left[ 1-\frac{1+w(t)}{\gamma k^2(1+3\tau H_0/Z)}\right]-\Lambda(t)=0. \label{22}
\end{equation}
Thus, if the parameter $w(t)$ is chosen as
\begin{equation}
w(t)=-1-\frac{k^2}{3H_0^2}Z(Z+3\tau H_0), \label{23}
\end{equation}
the "cosmological constant" becomes equal to
\begin{equation}
\Lambda(t)=1+\frac{3\gamma H_0^2}{Z^2}. \label{24}
\end{equation}
If $t\rightarrow +\infty$, then $w\rightarrow -\infty, ~\Lambda \rightarrow -\infty$, and the universe lies in the phantom region.

\bigskip

{\noindent 2) \bf A milder variant: the Little Rip scenario.}

\bigskip
Option 1) above was concerned with the Big Rip, meaning that the future singularity is encountered in a finite time. The Little Rip is a milder variant, as the time needed to obtain the singularity is infinite. Taking the equation of state in the form $p=-\rho-A\sqrt \rho$ with $A$ the same positive constant as in Eq.~(\ref{2}), we get \cite{brevik12B}
\begin{equation}
\rho=\frac{\xi_0^2}{9A^2}\left[ 1+\left(\frac{3A\sqrt{\rho_0}}{\xi_0}-1\right)\exp \left(\frac{1}{2}\sqrt{3}\,At\right)\right]^2, \label{25}
\end{equation}
which shows that the increase of $\rho$ towards infinity occurs only exponentially. We now find
\begin{equation}
\dot{\rho}=(3A\sqrt{\rho_0}-\xi_0)\exp \left(\frac{\sqrt 3}{2}kAt\right)H. \label{26}
\end{equation}
The energy conservation law takes takes the form
\begin{equation}
\sqrt{3}AH+k\left\{ [1+w(t)]\frac{3}{k^2}H^2-\frac{\xi_0}{3}+\Lambda(t)\right\}=0. \label{27}
\end{equation}
We solve this equation with respect to $\Lambda(t)$ and insert for the parameter $w(t)$ the expression
\begin{equation}
w(t)=-1-\frac{\delta k^2}{3H^2}, \label{28}
\end{equation}
with $\delta$ a positive constant. Then we obtain
\begin{equation}
\Lambda(t)=\frac{\xi_0}{3}+\delta-\frac{\sqrt 3}{k}AH. \label{29}
\end{equation}
In this case the LR is caused by the quantity $w$. When $t\rightarrow \infty$, $w(t)\rightarrow -1,~\Lambda(t)\rightarrow -\infty$. The future behavior of this universe will depend on the choice of the model parameters $\xi_0, A$ and $\delta$.

Consequently, if we start from a perfect fluid whose equation of state is given in the form (\ref{4}), within the framework of 4d FRW cosmology, we realize the viscous Little Rip via the choice (\ref{28}) for the parameter $w(t)$, corresponding to the expression (\ref{29}) for $\Lambda(t)$.

\section{A one-component dark fluid}

There is an alternative way of approaching the problem, namely to consider the cosmic fluid as a one-component fluid. The universe can be assumed to start from the present time $t=0$ as an ordinary viscous fluid with bulk viscosity $\zeta$, developing with time according to the Friedman equations in the viscous era towards a future singularity. We assume that the EoS parameter $w<-1$, so that the future singularity should on the basis of these conditions be unavoidable. Before the singularity is encountered we assume, however, that at some instant $t=t_*$ there occurs a sudden transition of the whole fluid into a turbulent state after which the EoS parameter is $w_{\rm turb}>-1$ and the pressure is equal to $p_{\rm turb}=w_{\rm turb}\,\rho_{\rm turb}$. On the laminar side of the transition point, $p_* =w\rho_* <0$, while on the turbulent side, $p_*=w_{\rm turb}\,\rho_*$ will even be positive if $w_{\rm turb}>0$. The density is continuous at $t=t_*$ whereas the pressure is not.
For simplicity we now take $\zeta, w$ and $w_{\rm turb}$ to be constants.

In the viscous era $0<t<t_*$ the energy density is \cite{brevik12B}
\begin{equation}
\rho=\frac{\rho_0\,e^{2t/t_c}}{[1-\frac{3}{2}|\gamma|H_0t_c(e^{t/t_c}-1)]^2}, \label{30}
\end{equation}
where $t_c$ is the "viscosity time"
\begin{equation}t_c=\left( \frac{3}{2}k^2\zeta\right)^{-1}. \label{31}
\end{equation}
From this we can calculate the energy density at $t=t_*$.

Take now the derivative of the energy density with respect to cosmic time,
\begin{equation}
\dot{\rho}=\frac{2\rho_0\,e^{2t/t_c}}{t_c}\,\frac{1+\frac{3}{2}|\gamma|H_0t_c}{[1-\frac{3}{2}|\gamma|H_0t_c(e^{t/t_c}-1)]^3}, \label{32}
\end{equation}
and use Eqs.~(\ref{4}), (\ref{5}), (\ref{30}) and (\ref{32}) to obtain the energy conservation law
\begin{equation}
\frac{2}{t_c}\left( 1+\frac{3}{2}|\gamma|H_0t_c\right) +3[1+w(t)]+k^2e^{\frac{t}{t_c}}\frac{\Lambda(t)}{H^2}=0. \label{33}
\end{equation}
When solving this with respect to $\Lambda(t)$,
\begin{equation}
\Lambda(t)=-\frac{H^2}{k^2e^{t/t_c}}\left\{ 3[1+w(t)]+\frac{2}{t_c}\left( 1+\frac{3}{2}|\gamma| H_0t_c\right)\right\} \label{34}
\end{equation}
we obtain, by choosing the parameter $w(t)$ in the form
\begin{equation}
w(t)=-1-\frac{\delta}{3H^2}e^{\frac{t}{t_c}} \label{35}
\end{equation}
with $\delta$ a positive constant, the following expression for $\Lambda(t)$:
\begin{equation}
\Lambda(t)=\frac{\delta}{k^2}-\frac{2\left( 1+\frac{3}{2}|\gamma| H_0t_c\right)}{k^2t_c}\,H^2e^{-\frac{t}{t_c}}. \label{36}
\end{equation}
If $t\rightarrow +\infty$, then $\Lambda(t)\rightarrow \delta/k^2$.

In the turbulent era we make in the expression (\ref{30}) the substitutions $t_c\rightarrow +\infty, t\rightarrow t-t_*, \gamma \rightarrow \gamma_{\rm turb}~(\gamma_{\rm turb}>0)$ and $\rho_0\rightarrow \rho_*$. Then \cite{brevik12B}
\begin{equation}
\rho=\frac{\rho_*}{\left[ 1+\frac{3}{2}\gamma_{\rm turb}H_*(t-t_*)\right]^2}, \label{37}
\end{equation}
and we now find
\begin{equation}
\dot{\rho}=-\frac{9}{k^2}\gamma_{\rm turb}H^3. \label{38}
\end{equation}
The energy conservation law is
\begin{equation}
\frac{3}{k^2}\gamma_{\rm turb}H^2-[1+w(t)]\frac{3}{k^2}H^2-\Lambda(t)=0. \label{39}
\end{equation}
Solving (\ref{39}) with respect to $\Lambda(t)$,
\begin{equation}
\Lambda(t)=\frac{3}{k^2}H^2[\gamma_{\rm turb}-w(t)-1], \label{40}
\end{equation}
and choosing the parameter $w(t)$ as
\begin{equation}
w(t)=-1-\frac{\delta k^2}{3H^2} \label{41}
\end{equation}
with $\delta$ a positive constant, we find the "cosmological constant" to be
\begin{equation}
\Lambda(t)=\delta +\frac{3\gamma_{\rm turb}}{k^2}H^2. \label{42}
\end{equation}
As both $\rho$ and $H$ go to zero when $t\rightarrow \infty$ in the turbulent era, it follows from the expression (\ref{42}) that $\Lambda(t)\rightarrow \delta$ when $t\rightarrow \infty$.

Thus, we have shown how the transition of a one-component cosmic fluid from the viscous era into the turbulent era can alternatively be looked upon as a 4d FRW cosmology situation in which the EoS equation takes the general form (\ref{4}) above.

\section{Conclusion}

This investigation can be taken as a demonstration of the diversity of cosmological fluid mechanical theory. Our starting point was the inclusion of turbulence in the cosmic fluid; this is natural approach in view of the fundamental property of classical fluids in general. As is known, the dark energy is often considered to be some kind of a classical fluid with unusual properties. It would seem physically reasonable to think that turbulence phenomena may be important for the dark energy, especially in the very late violent universe. What we have essentially shown, is that an equivalent description of viscous Little Rip cosmology for the dark fluid in the late universe can be obtained in terms of an inhomogeneous fluid within the framework of 4d FRW cosmology. The central form of the EoS equation is that of Eq.~(\ref{4}) above.

Thus a {\it two-component model}, in which the fluid system was assumed to be quasi-stationary with turbulent properties, was treated as an inhomogeneous fluid within 4d FRW cosmology.

A {\it one-component model} was also treated as an inhomogeneous fluid, in the viscous epoch as well as in the turbulent epoch, also in that case in terms of an inhomogeneous fluid in the 4d FRW theory.

The effect of turbulence and/or viscosity may thus always be tracked back to an appropriate effective equation of state.

\end{document}